\documentclass[epj,final,floats,nopacs]{svjour}
\usepackage{graphicx}
\usepackage{dcolumn}
\usepackage{bm}
\usepackage{color}
\usepackage{amsmath}
\usepackage{amssymb}
\usepackage{hyperref}

\usepackage{graphics}
\begin{document}
\title{Density functional theory versus quantum Monte Carlo simulations of Fermi gases in the optical-lattice arena}

\author{S. Pilati\inst{1,2} \and I. Zintchenko\inst{3} \and M. Troyer\inst{3,4} \and F. Ancilotto\inst{1,5}
}                     
\institute{Dipartimento di Fisica e Astronomia ``Galileo Galilei'', Universit\`a di Padova, via Marzolo 8, I-35131 Padova, Italy \and 
School of Science and Technology, Physics Division, University of Camerino, Via Madonna delle Carceri 9, I-62032 Camerino, Italy \and 
Theoretische Physik, ETH Zurich, 8093 Zurich, Switzerland \and Quantum Architectures and Computation Group, Microsoft Research, Redmond, WA (USA) \and CNR-IOM Democritos, via Bonomea 265, 34136 Trieste, Italy}

\date{}
%
\abstract{
We benchmark the ground state energies and the density profiles of atomic repulsive Fermi gases in
optical lattices computed via Density Functional Theory (DFT) against the results of diffusion Monte Carlo (DMC) simulations. 
The main focus is on a half-filled one-dimensional optical lattices, for which the DMC simulations performed within the fixed-node approach provide unbiased results.
This allows us to demonstrate that the local spin-density approximation (LSDA) to the
exchange-correlation functional of DFT is 
very accurate in the weak and intermediate interactions regime, 
and also to underline its limitations close to the strongly-interacting Tonks-Girardeau limit and in very deep optical lattices. 
We also consider a three dimensional optical lattice at quarter filling, 
showing also in this case the high accuracy of the LSDA in the moderate interaction regime.
The one-dimensional data provided in this study may represent a useful 
benchmark to further develop DFT methods beyond the LSDA and 
they will hopefully motivate experimental studies to accurately 
measure the equation of state of Fermi gases in higher-dimensional geometries.
}

\authorrunning{S. Pilati \emph{et al.}}
\titlerunning{Density functional theory versus quantum Monte Carlo simulations}

\maketitle
\section{Introduction}
\label{intro}
Kohn-Sham density functional theory (DFT) is the most widely used theoretical tool in material science
and in quantum chemistry \cite{burke2012perspective}. Its main ingredient is an accurate approximation
for the exchange-correlation energy-density functional for the electron system.
The basic approach consists in approximating 
this functional within the Local Density Approximation (LDA) or, in the case of spin-polarized systems,
the Local Spin-Density Approximation (LSDA), using as an input accurate results of
{\it ab initio}
calculations of the equation of state of the uniform electron system.
The LSDA allowed researchers to make quite accurate 
predictions for the ground state properties of a huge variety of materials \cite{ParrYang,KohnBecke}. 
Known limitations of the LSDA approach to the study of condensed matter systems 
are a less accurate representation of
excited-state properties (because the DFT is a ground state theory), 
and the partial neglect of strong 
electron-electron correlation effects in which electron-electron repulsion plays a prominent role,
like those
arising, e.g., between core electrons in transition-metals and transition-metal compounds.
While numerous, more sophisticated approximations than LSDA exist, 
including, e.g, generalized gradient approximations (GGA), meta-GGA, hyper-GGA, hybrid and generalized random-phase approximations\cite{perdew2001jacob,tao2003climbing,staroverov2004tests} or the LDA+$U$ methods \cite{anisimov1991band} including an effective Hubbard interaction term $U$, there is 
at present no systematic procedure to improve 
the accuracy of existing approximations 
to the DFT of electron systems,
and to systematically converge to the exact 
density functional.\\
%
%
%

In recent years, ultracold atoms trapped in optical lattices (OLs) have proven 
to be an ideal platform to perform quantum simulations of 
phenomena in the presence of strong inter-particle 
correlations \cite{bloch2008many}. Most of the early theoretical works are
focussed on single-band discrete-lattice Hamiltonians --- 
the most relevant being the Hubbard model --- which 
properly describe the experimental realization of such systems  
if the OL is very deep and the interactions are sufficiently weak~\cite{jaksch2005cold}. 
These models capture the phenomenology of strongly correlated systems, 
but they do not allow to make quantitative predictions of 
real materials' properties, as opposed to DFT methods. 
More recently, researchers addressed also shallow OLs employing 
continuous-space models, studying phenomena 
such as itinerant ferromagnetism~\cite{pilati2014},  bosonic superfluid-Mott insulator transitions~\cite{pilati2012}, and pinning localization transitions~\cite{de2012phase,boeris2016mott,astrakharchik2016one}.
It has also been proposed to use OL experiments as a testbed to develop more accurate approximations
for the exchange-correlation functional of DFT~\cite{dft}. In this respect, cold-atom systems offer
crucial advantages with respect to solid-state systems, since experimentalists 
are able to independently control the density inhomogeneity and the interaction strength  by tuning, respectively, the OL intensity and a magnetic field close to a Feshbach resonance.\\ 
%
%
DFT methods have already been employed to study ultracold fermionic gases, allowing to investigate phenomena such as  ferromagnetism and antiferromagnetism in repulsive Fermi gases in shallow OLs~\cite{dft}, vortex dynamics in superfluid Fermi gases~\cite{bulgac2011real,bulgac2014quantized}, superfluidity and density modulations in dipolar Fermi gases~\cite{ancilotto2016kohn},  vortices in rotating dipolar Fermi gases~\cite{ancilotto2015kohn}, and the formation of ferromagnetic domains in trapped clouds~\cite{zintchenko2016ferromagnetism}.
These studies employed exchange-correlation functionals based on the LSDA. However, it has not yet been analyzed in detail in which regimes this approximation is reliable.\\

The main goals of this article are (i) to assess the accuracy of the LSDA  for repulsive Fermi gases in
OLs and (ii) to provide an accurate benchmark for future studies aiming at developing beyond-LSDA approximations. To this aim, we mostly focus  on the one-dimensional geometry, for which quantum Monte Carlo (QMC) simulations based on the fixed-node method to circumvent the sign problem provide exact results~\cite{ceperley1991fermion}, within statistical uncertainties. Quantum fluctuations are known to play a more relevant role in one-dimension than in higher dimensional geometries, implying that the case we consider is a challenging testbed for the LSDA.
A systematic comparison between DFT calculations of ground state energies and density profiles for a
half-filled OL against the (exact) outcomes of the QMC simulations is presented. This allows us to
map the regime of OL intensities and interaction strengths 
where the LSDA is accurate.
Furthermore, we consider a three-dimensional repulsive Fermi gas in a simple-cubic OL at quarter
filling, making also in this case a comparison between DFT calculations and QMC simulations for the
ground state energies.\\

The rest of the article is organized as follows: Secs.~\ref{secdft} and \ref{secqmc} provide the main details of the DFT calculations and of the QMC simulations, respectively. In Sec~\ref{sec1D} the results for the ground state energy and the density profiles of a half-filled one-dimensional OL are discussed.  Section~\ref{sec3D} reports predictions for the ground state energy of the three-dimensional Fermi gas at quarter filling. Our conclusions and the outlook are reported in Sec.~\ref{conclusions}.

\begin{figure}
  \includegraphics[width=1.0\columnwidth]{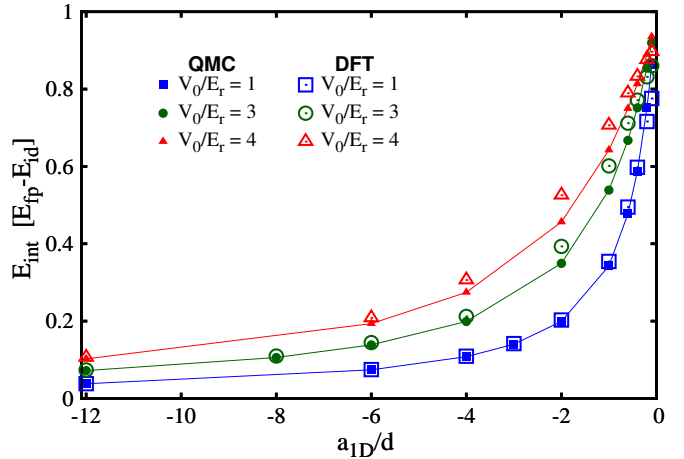}
\caption{(Color online). Ground state interaction energy $E_{\mathrm{int}}= E-E_{\mathrm{id}}$ of a one-dimensional repulsive Fermi gas $E$ in a half filled OL,  as a function of the interaction parameter $a_{1D}/d$, where $a_{1D}$ is the one-dimensional scattering length and $d$ is the OL periodicity. $E$, $E_{\mathrm{id}}$, and $E_{\mathrm{fp}}$ are the energies of the interacting, the noninteracting, and the fully-polarized (Tonks-Girardeau) gas, respectively. The three datasets correspond to three OL intensities $V_0$, expressed in unit of the recoil energy $E_r = \hbar^2\pi^2/(2md^2)$. Empty symbols represent DFT results, full symbols the QMC ones. Here and in the other figures for the one-dimensional OL the system size is $L=26d$.}
\label{fig1}       
\end{figure}

\begin{figure}
  \includegraphics[width=1.0\columnwidth]{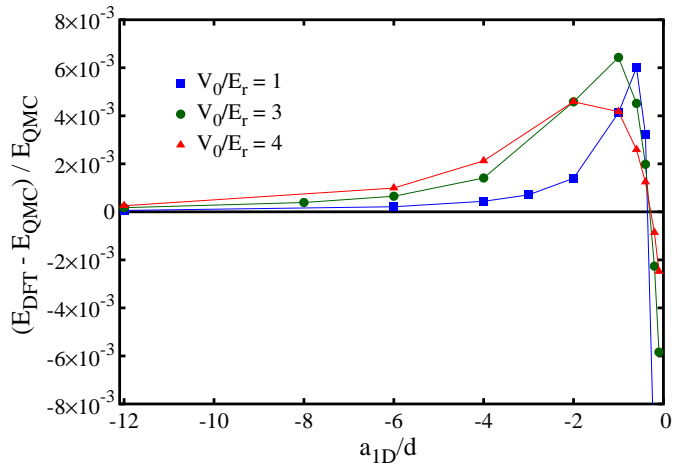}
\caption{Relative error of the DFT ground state energy $E_{\mathrm{DFT}}$ with respect to the exact QMC result $E_{\mathrm{QMC}}$ as a function of the interaction parameter $a_{1D}/d$. The density is fixed at half filling $n=1/d$, and the three datasets correspond to three OL intensities $V_0/E_r$.}
\label{fig2}       
\end{figure}


\section{Density functional theory for atomic Fermi gases in optical lattices}
\label{secdft}
In this Section, we consider a generic continuous-space  Hamiltonian describing a two-component Fermi gas in $D$ dimensions:
\begin{equation}
H = \sum_{\sigma=\uparrow,\downarrow}
       \sum_{i_\sigma=1             }^{N_\sigma      }\left(-\Lambda\nabla^2_{i_\sigma}  + V(\mathbf{x}_{i_\sigma}    )\right)
       +  \sum_{i_\uparrow,i_\downarrow}v(x_{i_\uparrow i_\downarrow}) 
       \;,
\label{hamiltonian}
\end{equation}
where $\Lambda=\hbar^2/2m$, with $m$ the atomic mass and $\hbar$ the reduced Planck constant. The indices $i_\uparrow$ and $i_\downarrow$ label atoms of the two components, which we refer to as spin-up and spin-down
fermions, respectively. 
The total number of fermions is $N = N_\uparrow + N_\downarrow$, and
 $x_{i_\uparrow i_\downarrow} = \left|\mathbf{x}_{i_\uparrow}-\mathbf{x}_{i_\downarrow}\right|$ is the relative distance between opposite-spin fermion pairs. 
$V(\mathbf{x})=V_0\sum_{\alpha=1}^D \sin^2\left(x_{\alpha}\pi/d\right)$ is a simple-cubic optical lattice potential with periodicity $d$ 
and intensity $V_0$, conventionally expressed in units of recoil energy $E_r=\Lambda\left(\pi/d\right)^2$. The system size $L$ is an integer multiple of the OL periodicity, and periodic boundary conditions are assumed.
$v(x)$ is a model repulsive potential, defined in Sections~\ref{sec1D} and \ref{sec3D} for the one-dimensional and the three-dimensional cases, respectively. Its intensity can be tuned in experiments using Feshbach resonances~\cite{chin}. Off-resonant intraspecies interactions in dilute atomic clouds are negligible at low temperature; hence they are not included in the Hamiltonian.\\

The Hohenberg-Kohn (HK) theorem \cite{hohenbergkohn}
states that the ground state energy $E$ of the many-body system
described by the Hamiltonian (\ref{hamiltonian})
is a functional of the one-particle densities $(\rho_\uparrow,\rho_\downarrow)$:

\begin{equation}
\label{spin_dependent_functional}
 E\left[\rho_\uparrow,\rho_\downarrow\right] =  \int d\mathbf{x}\,  V( \mathbf{x})
                       \left[ \rho_\uparrow\left(\mathbf{x}\right)  +  \rho_\downarrow\left(\mathbf{x}\right) \right]+
                       F\left[\rho_\uparrow,\rho_\downarrow \right].
\end{equation}
The first term is the potential energy due to the external potential $V(\mathbf{x})$. The second term
is an unknown but universal functional which includes the kinetic energy and interaction functionals, 
$F\left[\rho_\uparrow,\rho_\downarrow \right] \equiv T\left[\rho_\uparrow,\rho_\downarrow \right] +
V_{int}\left[\rho_\uparrow,\rho_\downarrow \right]
$,
but does not explicitly depend on the external potential $V( \mathbf{x})$. 

\begin{figure}
  \includegraphics[width=1.0\columnwidth]{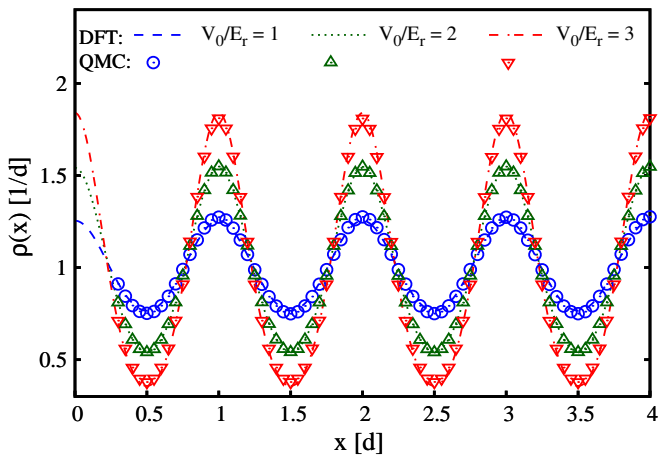}
\caption{Local density $\rho(x)$ of a repulsive Fermi gas in a half-filled one-dimensional OL, as a
function of the spatial coordinate $x$. The three datasets correspond to three OL intensities
$V_0/E_r$, while the interaction strength is fixed at the intermediate value $a_{1D}/d = -1$. $d$ is
the OL periodicity. The lines represent the DFT results, the empty symbols represent the QMC data. The
total system size is $L=26d$. Here and in Figs.~\ref{fig4}, \ref{fig5} and \ref{fig6} we only visualize 
the range $0\leqslant x \leqslant 4d$ for the sake of 
clarity. The QMC data for $x<0.3d$ have been removed to make the DFT curves more visible.}
\label{fig3}       
\end{figure}

\begin{figure}
  \includegraphics[width=1.0\columnwidth]{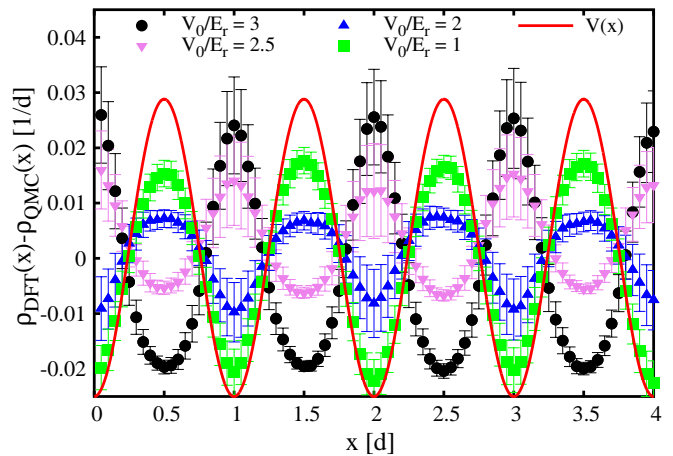}
\caption{Difference between the local density determined via DFT $\rho_{\mathrm{DFT}}(x)$ and the one obtained via QMC simulations $\rho_{\mathrm{QMC}}(x)$.
The interaction strength is $a_{1D}/d=-1$ and the (average) density is $n=1/d$.
The different symbols correspond to different OL intensities $V_0/E_r$. The thick continuous (red) curve represents the OL intensity profile, in arbitrary units.}
\label{fig4}       
\end{figure}

In the Kohn-Sham formulation of the HK theorem \cite{kohnsham} 
one writes the universal functional $F$ in the form

\begin{equation}
\label{kohn_sham_functional}
 F\left[\rho_\uparrow,\rho_\downarrow \right]=T_0\left[\rho_\uparrow,\rho_\downarrow \right]
+V_H\left[\rho_\uparrow,\rho_\downarrow \right]
+E_{XC}\left[\rho_\uparrow,\rho_\downarrow \right],
\end{equation}
where
$T_0$ is the kinetic energy of
a fictitious system of {\it non-interacting} fermions, with the same densities
of the original one,
described by single-particle orbitals
$\{ \phi _\uparrow  ^i({\mathbf x}),\,i=1,N_\uparrow \}$,
$\{ \phi _\downarrow  ^i({\mathbf x}),\,i=1,N_\downarrow \}$,
(such that the total density is simply 
$\rho ({\mathbf x})=\rho_\uparrow + \rho_\downarrow=\sum _i^{N_\uparrow }
|\phi _\uparrow ^i({\mathbf x})|^2+ \sum _i^{N_\downarrow }|\phi _\downarrow ^i({\mathbf x})|^2$~), 
$V_H\equiv 
{1 \over 2} \int d\mathbf{x}\, d\mathbf{x}^\prime \, \rho _\uparrow ({\mathbf x}) 
\rho _\downarrow ({\mathbf x}^\prime) v(x_{i_\uparrow i_\downarrow}) $
is the mean field (Hartree) expression for the interparticle interaction,
and 
$E_{XC} =(T-T_0)+(V_{int}-V_H)$ is the 
exchange-correlation energy functional.\\
The success of DFT for electrons (even at the LDA level of approximation)
is due to a partial cancellation between the 
terms contained in $E_{XC}$, thus reducing the impact
on the final results of any approximations done to approximate this term.
We will show that this holds true, at least for weak to intermediate 
interactions strengths, also for the fermionic gases investigated here.\\
While in the case of long-range Coulomb interactions, relevant for electrons in solids, 
one usually writes separately the mean-field energy $V_H$
and the exchange-correlation term $E_{XC}$, 
for short-range interactions relevant for atomic gases the mean-field term  depends only on the
local densities, and can thus be combined with the exchange-correlation term in 
a single energy functional $E_{HXC}$. 
For consistency with the literature, we will refer to it as exchange-correlation term
(instead of using the more appropriate ``Hartree-exchange-correlation'' name).\\
A simple yet often reliable treatment of $E_\mathrm{HXC}$ is the local spin-density approximation
\begin{equation}
\label{LSDAXC}
E_\mathrm{HXC}\left[\rho_\uparrow,\rho_\downarrow\right] = \int d\mathbf{x}\, \rho (\mathbf{x})
\epsilon_\mathrm{HXC} \left( \rho_\uparrow\left(\mathbf{x}\right) , \rho_\downarrow\left(\mathbf{x}\right) \right),
\end{equation}
where the functional is replaced by an integral over the interaction energy density 
of a uniform system with the same local spin-densities.
By imposing stationarity of the functional~(\ref{spin_dependent_functional}) with respect to
variations of the densities $\rho_\uparrow$ and $\rho_\downarrow$ one obtains a set of
Schr\"odinger-type equations
(the Kohn-Sham equations):
\begin{equation}
\label{kseq}
  \hat {H}_{\rm KS}\,\phi ^i_\sigma ({\bf x})\equiv \!
\left[-\Lambda \nabla ^2 \!+\! V({\bf x}) \!+\!
{\partial (\rho \epsilon _{HXC}) \over \partial \rho
_\sigma }\right] \!\phi ^i_\sigma ({\bf x})=
\epsilon _i \phi ^i_\sigma ({\bf x}).
\end{equation}%
From the eigenstates of the Kohn-Sham equations 
one can compute the density profiles and the ground state energy.\\

The LSDA exchange-correlation functional for one dimensional two-component Fermi gases with contact
interaction was derived in Ref.~\cite{abedinpour2007emergence} from the exact Bethe-Anstatz solution
for the ground state energy. The functional for three-dimensional Fermi gases with short-range
repulsive interactions has been obtained in Ref.~\cite{dft} using fixed-node diffusion Monte Carlo
simulations, similarly to the seminal work by Ceperley and Alder~\cite{ceperleyalder} who determined the equation of state of the uniform electron gas, upon which the parametrizations 
for $E_{XC}$ commonly employed in electronic-structure calculations have been built (see, e.g., ~\cite{perdew1986}).\\

\begin{figure}
  \includegraphics[width=1.0\columnwidth]{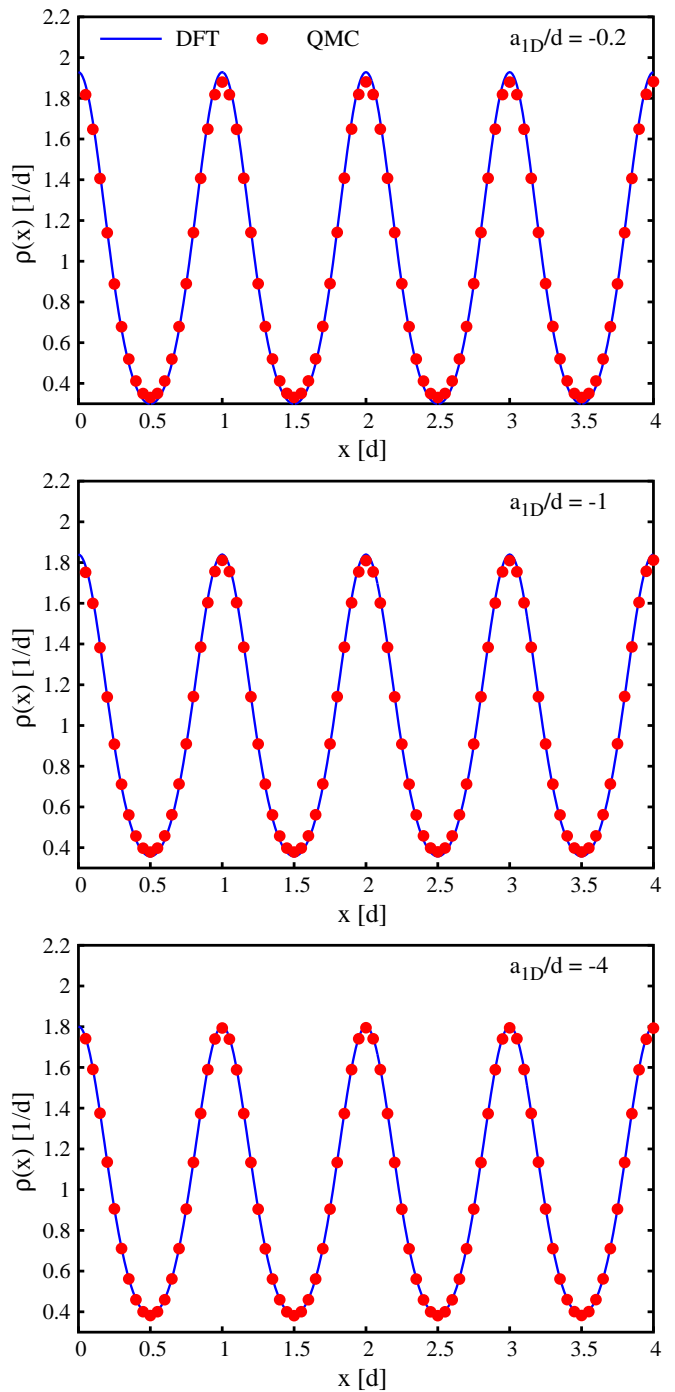}
\caption{Local density $\rho(x)$ of a repulsive Fermi gas in a half-filled one-dimensional OL, as a function of the spatial coordinate $x$. The three panels display data for three values of the interaction strength $a_{1D}/d$ at the same OL intensity $V_0/E_r=3$. The continuous lines represent the DFT results, the circles represent the QMC data.}
\label{fig5}
\end{figure}

\begin{figure}
  \includegraphics[width=1.0\columnwidth]{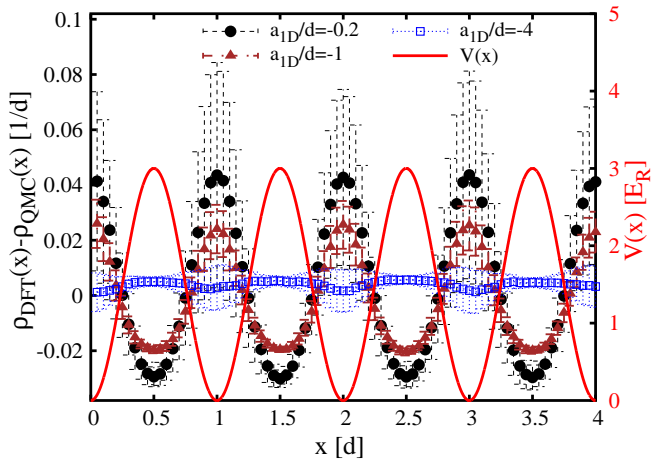}
\caption{Difference between the local density determined via DFT $\rho_{\mathrm{DFT}}(x)$ and the one obtained via QMC simulations $\rho_{\mathrm{QMC}}(x)$. The OL intensity is $V_0/E_r = 3$ and the (average) density is $n=1/d$. The different symbols correspond to different values of the interaction parameter $a_{1D}/d$. 
The think continuous (red) curve represent the OL intensity profile (referred to the right vertical axis).}
\label{fig6}
\end{figure}

\section{Fixed-node Diffusion Monte Carlo simulations}
\label{secqmc}
The ground state properties of the Hamiltonian (\ref{hamiltonian}) can be determined also via quantum Monte Carlo simulations based on the diffusion Monte Carlo (DMC) algorithm \cite{reynolds1982fixed}. 
The DMC algorithm allows one to sample the ground state wave function by stochastically 
evolving the Schr\"odinger equation in imaginary time. 
In order to circumvent the sign problem, which would otherwise hinder fermionic Monte Carlo simulations, one introduces the fixed-node constraint, which forces the nodal surface of the many-body wave function to coincide with that of a trial wave function $\psi_T$.
If the nodal surface of $\psi_T$ is exact, this method provides unbiased estimates of the ground state energy. In the general case, one obtains a rigorous upper bound for the exact ground state energy, which is very close to the exact result if the nodes of $\psi_T$ are good approximations of the ground state nodal surface (see, {\it e.g.}, \cite{foulkes}). 
In this study, we employ Jastrow-Slater trial wave functions defined as:
\begin{equation}
\psi_T({\bf X})= D_\uparrow(N_\uparrow) D_\downarrow(N_\downarrow) \prod_{i_\uparrow,i_\downarrow}f(x_{i_\uparrow i_\downarrow}) \;,
\label{psiT}
\end{equation}
where ${\bf X}=({\bf x}_1,..., {\bf x}_N)$ is the spatial configuration vector and $D_{\uparrow(\downarrow)}$ denotes the Slater determinant of single-particle orbitals of the particles with up (down) spin, and $x_{i_\uparrow i_\downarrow} =\left |\bf{x}_{i_\uparrow}- \bf{x}_{i_\downarrow} \right|$ indicates the relative distance between any opposite-spin fermion pair.
The Jastrow correlation term $f(x)$ is taken to be the solution of the s-wave radial Schr\"odinger equation describing the scattering of two particles in free space, as described in details in Refs. \cite{pilati2010,pilati2014}. Since $f(x)>0$, the nodal surface is determined by the Slater determinants, and therefore  by the choice for the single-particle orbitals.
We use the $N_\uparrow$ ($N_\downarrow$) lowest-energy single-particle eigenstates $\phi_{j}(\bf{x})$ (with $j=0,\dots,N_{\uparrow(\downarrow)}-1$), which satisfy the single-particle Schr\"odinger equation in the external potential: $\left[-\Lambda \nabla^2  +V({\bf x})\right]\phi_{j}(\mathbf{x}) = e_j \phi_{j}(\mathbf{x})$, with the eigenvalues $e_j$. We determine these orbitals via exact  diagonalization of the finite matrix obtained within a discrete-variable representation based on high-order finite-difference formulas for the Laplacian. The discretization error can be reduced at the point to be  negligible compared to the statistical uncertainty of the Monte Carlo simulation.\\

While the fixed-node constraint might introduce a systematic bias, the predictions made with this approach  have often been found to be extremely accurate. For example, the recent measurements performed at LENS with a strongly-repulsive Fermi gas in the upper branch of a Feshbach resonance\cite{valtolina2017exploring} --- which have been analyzed within the spin-fluctuation theory of Ref. \cite{recati2011spin} --- have been found to agree with previous predictions for the equation of state and the Stoner ferromagnetic instability obtained via fixed-node DMC simulations in Ref. \cite{pilati2010}.\\
Interestingly,  it was shown in Ref. \cite{ceperley1991fermion} that in the one-dimensional case the fixed-node approach is, in fact, exact  since  the nodal surface consists only of the many-particle configurations where two identical fermions occupy the same point. This implies that any Slater-determinant wave function, as the trial wave function we use in this article,  has the same nodes as the exact ground state~\cite{casula2008quantum,astrakharchikgiorgini,matveeva2016one}.
Therefore, the data we provide for the one-dimensional Fermi gas in the OL represent an exact benchmark, useful to measure the accuracy of the DFT calculations based on the LSDA or of any other computational tool.
 Furthermore, in order to compute unbiased expectation values also for the density profiles, we employ the standard forward walking technique~\cite{boronat}.

\section{One-dimensional atomic Fermi gas in an optical lattice}
\label{sec1D}
Let's consider a one-dimensional Fermi gas with a zero-range repulsive interaction defined as $v(x_{i_\uparrow} -x_{i_\downarrow})= g\delta(x_{i_\uparrow} -x_{i_\downarrow})$, where the coupling constant $g$ is related to the one-dimensional scattering length $a_{1D}$  by the relation: $g=-2\hbar^2/(ma_{1D})$. We address the case of repulsive interaction, where $g \geqslant0$ and, correspondingly $a_{1D} \leqslant 0$.
The  $a_{1D}\rightarrow -\infty $ limit corresponds to the noninteracting Fermi gas, while the $a_{1D}\rightarrow 0^- $ corresponds to the strongly-interacting limit  where distinguishable fermions fermionize \cite{girardeau1960,girardeau,astrakharchikgiorgini,jochim}, 
i.e. their energy and density profiles correspond to those of indistinguishable (i.e. spin polarized) fermions \cite{guan}. For consistency with the more familiar case of infinitely repulsive bosons,
 we refer to this limit as 
 Tonks-Girardeau limit~\cite{guanshu}.
 In the following, we parametrize the interaction strength with the adimensional ratio $a_{1D}/d$, where $d$ is the OL periodicity.\\
This one-dimensional model is relevant to describe the experimental setup of an ultracold atomic gas confined in a tight cigar-shaped waveguide, sufficiently strong to prevent thermal excitations to higher radial modes. In this regime, the values of the one-dimensional scattering length can be determined from the experimental parameters, specifically from the three-dimensional s-wave scattering length $a$ and the radial harmonic confining frequency~\cite{olshanii1998atomic}. The one-dimensional scattering length can be tuned from the noninteracting to the Tonks-Girardeau limit --- and also beyond --- by approaching a confinement induced resonance. This is can be performed by exploiting a Feshbach resonance to modify the three-dimensional scattering length and/or by tuning the strength of the radial confinement.\\

We focus on a half-filled OL at the density $n=1/d$, so that on average there is one fermion per well. In this configuration the correlation effects are enhanced and strong quasi long-range antiferromagnetic order arises as one increases the OL intensity $V_0$ and/or the interaction strength \cite{PhysRevA.96.021601}. Therefore this regime represents a challenging testbed for the DFT calculations performed within the LSDA. It is worth pointing out that  as a consequence of the Mermin-Wagner theorem in one dimension proper long-range antiferromagnetic order is not possible and the ground state is paramagnetic.\\
In Fig.~\ref{fig1} the ground state interaction energy of the half-filled OL is reported as a function of the interaction parameter $a_{1D}/d$, for three OL intensities. 
The numerical values corresponding to (a selection of) these datasets are reported in Table I of the supplemental material~\cite{suppmat}.
In the noninteracting $a_{1D}/d\rightarrow -\infty$ limit the ground state energy $E$ converges to the noninteracting gas results $E_{\mathrm{id}}$, so that the interaction energy defined as $E_{\mathrm{int}}=E-E_{\mathrm{id}}$ vanishes. In the strongly-interacting $a_{1D}/d \rightarrow 0$ limit, $E$ approaches the energy of a fully polarized (i.e., with $N_{\uparrow}=N$ and $N_{\downarrow}=0$) gas $E_{\mathrm{fp}}$, analogously to the case of bosons with 
infinitely repulsive delta-function interaction described by the Tonks-Girardeau theory~\cite{girardeau1960}.
The discrepancies between the DFT prediction and the exact QMC results are surprisingly small, in particular for the shallow lattice of intensity $V_0/E_r=1$. In order to better visualize these discrepancies, we display in Fig.~\ref{fig2} the relative error of the DFT prediction $E_{\mathrm{DFT}}$ with respect to the corresponding QMC result $E_{\mathrm{QMC}}$. One observes that this relative error is smaller than $1\%$ in a broad range of interaction strengths, up to $a_{1D}/d<-0.2$. Only very close to the Tonks-Girardeau limit  one has quite large (negative) relative errors.
It is also worth noticing that the relative error is non-monotonic, being positive for weak and intermediate interaction strengths and 
negative close to the Tonks-Girardeau limit.
\\

In order to shed light on the origin of the inaccuracy of the DFT prediction for the ground state
energy, it is useful to inspect also the predictions for the density profiles, which are 
one of the main ingredients of the DFT formalism. Figure~\ref{fig3} shows the total density
$\rho(x)=\rho_{\uparrow}(x)+\rho_{\downarrow}(x)$ as a function of the spatial coordinate $x$ at the
intermediate interaction strength $a_{1D}/d=-1$. 
The corresponding numerical values are reported in Table II of the supplemental on-line material~\cite{suppmat}. One notices again a remarkable level of accuracy,
at all three OL intensities considered. 
In order to highlight the (small) discrepancies, we show in  Fig.~\ref{fig4} the difference 
between the density profiles predicted by DFT and by QMC simulations. In shallow OLs with intensities $V_0/E_r \lesssim 2$ the DFT predictions exceed 
the QMC results at the peaks of the OL, meaning that DFT underestimates the density inhomogeneity.
Instead, in relatively deep OLs $V_0/E_r \gtrsim 2$, the DFT result is higher than the QMC one at the minima of the OL potential, meaning that in this case
 DFT overestimates the density  inhomogeneity.\\
Next, in Fig.~\ref{fig5} we show three density profiles
for three values of the interaction parameter $a_{1D}/d$, for a relatively deep OL with intensity
$V_0/E_r=3$. While for weak and moderately strong interactions the agreement between DFT and QMC
simulations is, again, remarkably accurate, at the strongest interaction strength $a_{1D}/d=-0.2$ the discrepancy becomes sizable. In order to better visualize this discrepancy, we plot in Fig.~\ref{fig6} the difference between the DFT and the QMC results. Consistently with the results discussed above, in this relatively deep OL the DFT prediction is higher than the exact QMC data at the minima of the OL potential, while it is lower than that at the maxima. 
%
%
Note that DFT applied to strongly localized electron systems also tends to favor, within the
LSDA, electron densities that are more inhomogeneous than the true ones, leading to 
overbinding of atoms in molecules, and overestimating
the calculated cohesive energies in solids. 
This well known deficiency of the LSDA for electrons is 
alleviated to a large extent by the use of 
the GGA approach, where functionals depend on the local density as well as
on the spatial variation of the density, $\nabla \rho$. Computationally such corrections are 
as simple to use as the LSDA itself.
This suggests
that a possible improvement over LSDA for fermionic gases could be the addition of 
gradient corrections $\alpha \nabla \rho$, with adjustable phenomenological 
parameters $\alpha $, to
the LSDA functional.

\begin{figure}
  \includegraphics[width=1.0\columnwidth]{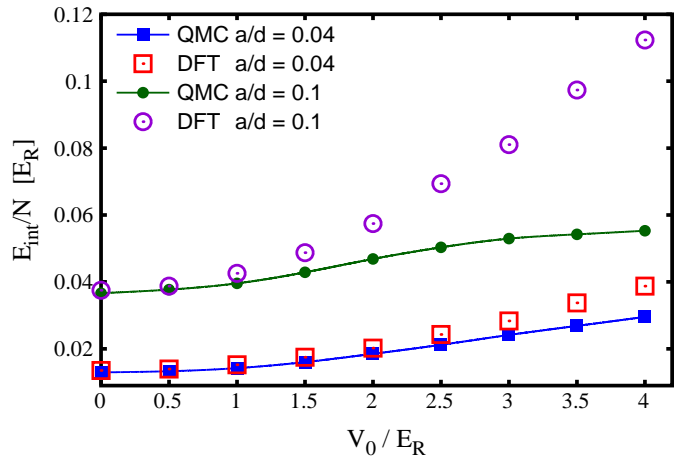}
\caption{Ground state interaction energy per particle $E_{\mathrm{int}}/N=(E-E_{\mathrm{ni}})/N$ of a repulsive Fermi gas in a three-dimensional OL at quarter filling $nd^3=0.5$, as a function of the OL intensity $V_0/E_r$. The energy unit is the recoil energy $E_r$. Data for two values of the interactions strength $a/d$ are shown. Full symbols connected by lines represent the QMC data, empty symbols represent the DFT results. The system size is $L=4d$.
}
\label{fig7}       
\end{figure}

\section{Three-dimensional atomic Fermi gas in a simple-cubic optical lattice}
\label{sec3D}
We now address a three-dimensional atomic Fermi gas in a simple-cubic OL. Since the zero-range (Fermi-Huang) pseudopotential supports two-body bound states in three-dimensions~\cite{huang}, we model the inter-species interactions using a purely repulsive potential with short but finite range, namely the hard-sphere model defined as: $v(r)=+\infty$ if $r<R_0$ and zero
otherwise.  This allows us to employ ground state computational methods, such as the DFT and the QMC methods considered in this article, while with the zero-range model the repulsive atomic state would be a highly excited (metastable) state, being the zero-temperature state a gas of bosonic molecules.
At zero temperature, the properties of a sufficiently dilute atomic
gas are universal, meaning that they depend only on the two-body scattering length $a$. For the hard-sphere model, one has $a=R_0$.
As the gas parameter $na^3$ increases, other details of the potential might become
relevant, the most important being the effective range and the $p$-wave scattering length. The regime where these nonuniversal effects become sizable has been carefully analyzed both in homogeneous gases~\cite{pilati2010} and in OL systems~\cite{pilati2014}. In this article, we consider a range of gas parameter where these effects are negligible.\\

We perform DFT calculations within the LSDA, using the exchange-correlation functional for the two-component Fermi gas with short-range interspecies interactions that has been reported in Ref.~\cite{dft}. This functional was obtained from  fixed-node DMC simulations of the equation of state of the homogeneous Fermi gas, and an accurate  parametrization based on the Fermi liquid and the polaron theories was provided.\\

We consider an OL at quarter filling, i.e., $nd^3=0.5$. Figure~\ref{fig7} displays the comparison between the DFT and the QMC results for the ground state energy as a function of the OL intensity 
\footnote{The QMC simulations have been performed with two system sizes, namely $L=4d$ and $L=6d$. We verified that including the finite-size correction corresponding to the noninteracting gas, as discussed in Ref.~\cite{LinZong}, the finite-size effect on the ground state energy becomes negligible.}.
In order to better visualize the discrepancies, the quantity plotted is the interaction energy $E_{\mathrm{int}}=E-E_{\mathrm{id}}$. Two interaction strengths are considered, corresponding to two values of the ratio $a/d$. At the moderate interaction strength $a/d=0.04$, the discrepancy is remarkably small even is relatively deep OLs. The relative error on the total energy $E$ reaches $0.2\%$ at $V_0/E_r=4$, which is the deepest lattice we consider. 
For very strong interactions $a/d = 0.1$, the DFT results agree with the QMC data only in shallow lattices $V_0/E_r\lesssim 1$, while significant discrepancies develop in deeper OLs. For example, at $V_0/E_r=4$ the relative error on the total energy $E$ is $1\%$.
This analysis shows how the interplay between the correlations induced by strong interatomic interactions and the pronounced inhomogeneity due to deep external potential causes the breakdown of the LSDA, meaning that more accurate approximations for the exchange-correlation functional are needed.\\
We emphasize that, in contrast with the one-dimensional case where the fixed-node DMC simulations
provide unbiased data, in three-dimensions the fixed-node DMC results represent an upper-bound for
the exact ground state energy. This upper-bound is believed to be extremely close to the exact
results, as demonstrated, for example,  by the agreement between fixed-node DMC simulations (in deep
OLs) with state-of-the-art constraint-path simulations of the Hubbard model~\cite{zhang,pilati2014},
and also with the recent experiments performed at LENS for strongly-repulsive homogeneous Fermi
gases mentioned in Sec.~\ref{secqmc}. Still, accurate experimental measurements of the zero-temperature equation of state in three-dimensional OLs would represent an extremely valuable benchmark for the fixed-node approach and for the DFT calculations.

\section{Conclusions and outlook}
\label{conclusions}
We performed a detailed benchmark of DFT 
calculations based on the LSDA for the exchange-correlation functional against QMC simulations. We
considered a one-dimensional Fermi gas in a half-filled  OL and a three-dimensional Fermi gas in a
simple-cubic OL at quarter filling. The one-dimensional case is special since the QMC results obtained
with the fixed-node approach are unbiased, being affected only by statistical uncertainties. This
allowed us to demonstrate that the LSDA is extremely accurate in a broad range of interaction
strengths and of OL intensities. Still, important inaccuracies of the LSDA emerge in the close vicinity of the
strongly-interacting Tonks-Girardeau limit and in very deep OLs (if interactions are not weak). 
We argue that the data we provide (see the supplemental material~\cite{suppmat}) 
represent an ideal testbed to further develop the Kohn-Sham DFT formalism beyond the LSDA. This
might include the use of gradient-dependent correction terms in the
total energy density functional.
Also in the case of the three-dimensional Fermi gas the agreement between DFT and QMC data is remarkable, at least in shallow OLs, and we hope that our study will motivate further experiments aiming at accurately measuring the equation of state and the density profiles of repulsive Fermi gas in OLs, in particular close to half-filling, a regime which is particularly challenging for any computational technique, including the QMC and the DFT methods we considered in this article. These measurements could be used as a testbed to develop more accurate exchange-correlation functional in higher dimensions.\\
The Kohn-Sham DFT formalism provides theoreticians with a useful computational tool to predict the properties of ultracold atomic gases in complex experimental configurations. Albeit approximate
in its practical implementations,
this method allows one to address much larger system sizes than those amenable to other computational methods such as, e.g., QMC algorithms, at the point to consider models of trapped clouds with realistic system sizes~\cite{zintchenko2016ferromagnetism}. Furthermore, it can be easily extended to simulate dynamical properties. The analysis we provide in this article is valuable since it maps out the regime where the most commonly adopted approximation for the exchange-correlation functional, namely the  LSDA, is reliable, thus providing a useful guide for future studies.\\

\noindent
We acknowledge the CINECA award under the ISCRA initiative, for the availability of high performance computing resources and support.
S. P. and F. A. acknowledge financial support from the BIRD 2016 project ``Superfluid properties of Fermi gases in optical potentials'' of the University of Padova.

\bibliographystyle{epj}
\bibliography{Ref}{}

\begin{thebibliography}{47}

\bibitem{burke2012perspective}
K.~Burke, J. Chem. Phys. \textbf{136}, 150901 (2012)

\bibitem{ParrYang}
R.G. Parr, W.~Yang, \emph{Density-Functional Theory of Atoms and Molecules}
  (Oxford University Press, New York, 1989)

\bibitem{KohnBecke}
W.~Kohn, A.D. Becke, R.G. Parr, J. Phys. Chem. \textbf{100}, 12974 (1996)

\bibitem{perdew2001jacob}
J.P. Perdew, K.~Schmidt, \emph{Jacob's ladder of density functional
  approximations for the exchange-correlation energy}, in \emph{AIP Conference
  Proceedings} (AIP, 2001), Vol. 577, pp. 1--20

\bibitem{tao2003climbing}
J.~Tao, J.P. Perdew, V.N. Staroverov, G.E. Scuseria, Phys. Rev. Lett.
  \textbf{91}, 146401 (2003)

\bibitem{staroverov2004tests}
V.N. Staroverov, G.E. Scuseria, J.~Tao, J.P. Perdew, Phys. Rev. B \textbf{69},
  075102 (2004)

\bibitem{anisimov1991band}
V.I. Anisimov, J.~Zaanen, O.K. Andersen, Phys. Rev. B \textbf{44}, 943 (1991)

\bibitem{bloch2008many}
I.~Bloch, J.~Dalibard, W.~Zwerger, Rev. Mod. Phys. \textbf{80}, 885 (2008)

\bibitem{jaksch2005cold}
D.~Jaksch, P.~Zoller, Ann. Phys. (N.Y.) \textbf{315}, 52 (2005)

\bibitem{pilati2014}
S.~Pilati, I.~Zintchenko, M.~Troyer, Phys. Rev. Lett. \textbf{112}, 015301
  (2014)

\bibitem{pilati2012}
S.~Pilati, M.~Troyer, Phys. Rev. Lett. \textbf{108}, 155301 (2012)

\bibitem{de2012phase}
F.~De~Soto, M.~Gordillo, Phys. Rev. A \textbf{85}, 013607 (2012)

\bibitem{boeris2016mott}
G.~Bo{\'e}ris, L.~Gori, M.D. Hoogerland, A.~Kumar, E.~Lucioni, L.~Tanzi,
  M.~Inguscio, T.~Giamarchi, C.~D'Errico, G.~Carleo et~al., Phys. Rev. A
  \textbf{93}, 011601 (2016)

\bibitem{astrakharchik2016one}
G.E. Astrakharchik, K.V. Krutitsky, M.~Lewenstein, F.~Mazzanti, Phys. Rev. A
  \textbf{93}, 021605 (2016)

\bibitem{dft}
P.N. Ma, S.~Pilati, M.~Troyer, X.~Dai, Nat. Phys. \textbf{8}, 601 (2012)

\bibitem{bulgac2011real}
A.~Bulgac, Y.L. Luo, P.~Magierski, K.J. Roche, Y.~Yu, Science \textbf{332},
  1288 (2011)

\bibitem{bulgac2014quantized}
A.~Bulgac, M.M. Forbes, M.M. Kelley, K.J. Roche, G.~Wlaz{\l}owski, Physical
  review letters \textbf{112}, 025301 (2014)

\bibitem{ancilotto2016kohn}
F.~Ancilotto, Phys. Rev. A \textbf{93}, 053627 (2016)

\bibitem{ancilotto2015kohn}
F.~Ancilotto, Phys. Rev. A \textbf{92}, 061602 (2015)

\bibitem{zintchenko2016ferromagnetism}
I.~Zintchenko, L.~Wang, M.~Troyer, Eur. Phys. J. B \textbf{89}, 180 (2016)

\bibitem{ceperley1991fermion}
D.M. Ceperley, J. Stat. Phys. \textbf{63}, 1237 (1991)

\bibitem{chin}
C.~Chin, R.~Grimm, P.~Julienne, E.~Tiesinga, Rev. Mod. Phys. \textbf{82}, 1225
  (2010)

\bibitem{hohenbergkohn}
P.~Hohenberg, W.~Kohn, Phys. Rev. \textbf{136}, B864 (1964)

\bibitem{kohnsham}
W.~Kohn, L.J. Sham, Phys. Rev. \textbf{140}, A1133 (1965)

\bibitem{abedinpour2007emergence}
S.H. Abedinpour, M.~Polini, G.~Xianlong, M.P. Tosi, Phys. Rev. A \textbf{75},
  015602 (2007)

\bibitem{ceperleyalder}
D.M. Ceperley, B.J. Alder, Phys. Rev. Lett. \textbf{45}, 566 (1980)

\bibitem{perdew1986}
J.P. Perdew, W.~Yue, Phys. Rev. B \textbf{33}, 8800 (1986)

\bibitem{reynolds1982fixed}
P.J. Reynolds, D.M. Ceperley, B.J. Alder, W.A. Lester~Jr, J. Chem. Phys.
  \textbf{77}, 5593 (1982)

\bibitem{foulkes}
W.~Foulkes, L.~Mitas, R.~Needs, G.~Rajagopal, Rev. Mod. Phys. \textbf{73}, 33
  (2001)

\bibitem{pilati2010}
S.~Pilati, G.~Bertaina, S.~Giorgini, M.~Troyer, Phys. Rev. Lett. \textbf{105},
  030405 (2010)

\bibitem{valtolina2017exploring}
G.~Valtolina, F.~Scazza, A.~Amico, A.~Burchianti, A.~Recati, T.~Enss,
  M.~Inguscio, M.~Zaccanti, G.~Roati, Nature Physics \textbf{13}, 704 (2017)

\bibitem{recati2011spin}
A.~Recati, S.~Stringari, Phys. Rev. Lett. \textbf{106}, 080402 (2011)

\bibitem{casula2008quantum}
M.~Casula, D.~Ceperley, E.J. Mueller, Phys. Rev. A \textbf{78}, 033607 (2008)

\bibitem{astrakharchikgiorgini}
G.E. Astrakharchik, D.~Blume, S.~Giorgini, L.P. Pitaevskii, Phys. Rev. Lett.
  \textbf{93}, 050402 (2004)

\bibitem{matveeva2016one}
N.~Matveeva, G.E. Astrakharchik, New J. Phys. \textbf{18}, 065009 (2016)

\bibitem{boronat}
J.~Boronat, in \emph{Microscopic Approaches to Quantum Liquids in Confined
  Geometries}, edited by E.~Krotscheck, J.~Navarro (World Scientific,
  Singapore, 2002, 2002), chap.~2, pp. 21--90

\bibitem{girardeau1960}
M.~Girardeau, J. Math. Phys. \textbf{1}, 516 (1960)

\bibitem{girardeau}
M.D. Girardeau, Phys. Rev. A \textbf{82}, 011607 (2010)

\bibitem{jochim}
G.~Z{\"u}rn, F.~Serwane, T.~Lompe, A.N. Wenz, M.G. Ries, J.E. Bohn, S.~Jochim,
  Phys. Rev. Lett. \textbf{108}, 075303 (2012)

\bibitem{guan}
L.~Guan, S.~Chen, Y.~Wang, Z.Q. Ma, Phys. Rev. Lett. \textbf{102}, 160402
  (2009)

\bibitem{guanshu}
L.~Guan, S.~Chen, Phys. Rev. Lett. \textbf{105}, 175301 (2010)

\bibitem{olshanii1998atomic}
M.~Olshanii, Phys. Rev. Lett. \textbf{81}, 938 (1998)

\bibitem{PhysRevA.96.021601}
S.~Pilati, L.~Barbiero, R.~Fazio, L.~Dell'Anna, Phys. Rev. A \textbf{96},
  021601 (2017)

\bibitem{suppmat}
See supplemental on-line material.

\bibitem{huang}
K.~Huang, C.N. Yang, Phys. Rev. \textbf{105}, 767 (1957)

\bibitem{LinZong}
C.~Lin, F.H. Zong, D.M. Ceperley, Phys. Rev. E \textbf{64}, 016702 (2001)

\bibitem{zhang}
C.C. Chang, S.~Zhang, D.M. Ceperley, Phys. Rev. A \textbf{82}, 061603 (2010)

\end{thebibliography}

\cleardoublepage

\begin{center}

{{\large \bf Supplemental material for ``Density functional theory versus quantum Monte Carlo simulations of Fermi gases in the optical-lattice arena''}}
\end{center}

In Table~\ref{table1} we report the ground-state energy per particle $E/N$ (in units of  $\hbar^2/(2md^2)$) of a one-dimensional Fermi gas in an optical lattice (OL) computed via fixed-node diffusion Monte Carlo simulations, as a function of the interaction parameter $a_{1D}/d$, where $a_{1D}$ is the one-dimensional scattering length and $d$ is the optical-lattice periodicity. Two OL intensities $V_0/E_r$ are considered, where $E_{r}=\hbar^2\pi^2/(2md^2)$ is the recoil energy, with $\hbar$ the reduced Planck constant and $m$ the particle mass. The system size is $L=26d$, and the particle number is $N=N_{\uparrow}+N_{\downarrow}=26$, with $N_{\uparrow}=13$ spin-up  and $N_{\downarrow}=13$  spin-down fermions. Periodic boundary conditions are assumed. 
The first and the last rows correspond, respectively, to the energies of the noninteracting gas $a_{1D}/d=-\infty$ and of the fully polarized gas (i.e., with $N_{\uparrow}=26$ and $N_{\downarrow}=0$), which is reached in the Tonks-Girardeau limit $a_{1D}/d\rightarrow 0$. These energies have been computed via exact numerical diagonalization.
In the thermodynamic limit, the ideal-gas energy per particle 
at $V_0/E_r=1$ is $E/N\cong5.42151 \hbar^2/(2md^2)$, meaning that finite size effects are below $0.1\%$. By performing diffusion Monte Carlo simulations with systems sizes up to $L=54d$ we verified that in the interacting case the finite size effect is of the same order of magnitude.\\
\begin{table}[h]
\centering
\begin{tabular}{| c | c | c |}
\hline
$a_{1D}/d$ & $E/N$ for  $V_0/E_r=1 $ & $E/N$ for $V_0/E_r=4$  \\
\hline
$-\infty$ & $5.41691$ & $15.751888$ \\
$-16$ & $5.48029(2)$ & $15.83557(3)$ \\
$-12$ & $5.50076(3)$ & $15.86193(4)$  \\
$-6$  & $5.57983(3)$ & $15.95982(7)$  \\
$-4$  & $5.65446(7)$ & $16.04662(8)$  \\
$-2$  & $5.8548(1)$ & $16.2419(4)$  \\
$-1$  & $6.1700(2)$ & $16.4417(5)$ \\
$-0.6$ & $6.4656(2)$ & $16.5568(6)$ \\
$-0.4$ & $6.7087(2)$ & $16.6254(5)$ \\
$-0.2$ & $7.0689(4)$ & $16.7070(7)$ \\
$-0.1$ & $7.3126(5)$ & $16.7563(7)$ \\
$0$ & $7.6146225$ & $16.825781$ \\
\hline    
\end{tabular}
\caption{Energy per particle $E/N$ for a one-dimensional Fermi gas in a half filled optical lattice as a function of the interaction parameter $a_{1D}/d$. Two optical lattice intensities $V_0/E_r$ are considered.}
\label{table1}
\end{table}
\newpage

In Table~\ref{table2} we report the local density $\rho(x)$ (in units of $1/d$)  as a function of the spatial coordinate $x$ (in units of $d$), for three combinations of OL intensity  $V_0/E_r$ and interaction parameter $a_{1D}/d$. The total system size and the particle numbers are as in Table I. Since within statistical uncertainties the density profile has periodicity $d$, we report only data corresponding to the first well of the optical lattice $x\in\left[0:d\right]$.
\begin{table}[h]
\begin{tabular}{| c | c |  c | c |}
\hline
$x$ & & & \\
\hline
$0.05$&   $	1.741 \pm	0.007$   &      $	1.812\pm 	0.027$   &      $	1.259\pm 	0.004$ \\
$0.10$&   $	1.591 \pm	0.006$   &      $	1.642\pm 	0.024$   &      $	1.216\pm 	0.004$ \\
$0.15$&   $	1.375 \pm	0.005$   &      $	1.402\pm 	0.020$   &      $	1.151\pm 	0.004$ \\
$0.20$&   $	1.135 \pm	0.004$   &      $	1.136\pm 	0.016$   &      $	1.072\pm 	0.003$ \\
$0.25$&   $	0.906 \pm	0.004$   &      $	0.885\pm 	0.012$   &      $	0.989\pm 	0.003$ \\
$0.30$&   $	0.711 \pm	0.003$   &      $	0.676\pm 	0.008$   &      $	0.911\pm 	0.003$ \\
$0.35$&   $	0.561 \pm	0.002$   &      $	0.518\pm 	0.006$   &      $	0.844\pm 	0.003$ \\
$0.40$&   $	0.459 \pm	0.002$   &      $	0.411\pm 	0.004$   &      $	0.794\pm 	0.002$ \\
$0.45$&   $	0.400 \pm	0.002$   &      $	0.349\pm 	0.003$   &      $	0.762\pm 	0.002$ \\
$0.50$&   $	0.381 \pm	0.002$   &      $	0.330\pm 	0.002$   &      $	0.752\pm 	0.002$ \\
$0.55$&   $	0.400 \pm	0.002$   &      $	0.350\pm 	0.003$   &      $	0.762\pm 	0.002$ \\
$0.60$&   $	0.459 \pm	0.002$   &      $	0.411\pm 	0.004$   &      $	0.793\pm 	0.003$ \\
$0.65$&   $	0.561 \pm	0.002$   &      $	0.518\pm 	0.005$   &      $	0.844\pm 	0.003$ \\
$0.70$&   $	0.710 \pm	0.003$   &      $	0.676\pm 	0.008$   &      $	0.911\pm 	0.003$ \\
$0.75$&   $	0.905 \pm	0.004$   &      $	0.886\pm 	0.011$   &      $	0.990\pm 	0.003$ \\
$0.80$&   $	1.134 \pm	0.005$   &      $	1.136\pm 	0.015$   &      $	1.072\pm 	0.003$ \\
$0.85$&   $	1.374 \pm	0.006$   &      $	1.402\pm 	0.019$   &      $	1.150\pm 	0.004$ \\
$0.90$&   $	1.589 \pm	0.007$   &      $	1.643\pm 	0.023$   &      $	1.215\pm 	0.004$ \\
$0.95$&   $	1.740 \pm	0.008$   &      $	1.812\pm 	0.026$   &      $	1.258\pm 	0.004$ \\
$1.00$&   $	1.794 \pm	0.008$   &      $	1.874\pm 	0.027$   &      $	1.274\pm 	0.004$ \\
\hline    
\end{tabular}
\caption{Density profile $\rho(x)$ as a function of the spatial coordinate $x$. 
The second and the third columns correspond to the interaction parameters $a_{1D}/d=-4 $ and $a_{1D}/d=-0.2$, respectively, and to 
the OL intensity $V_0/E_r=3$. The fourth column corresponds to  $a_{1D}/d=-1 $ and $V_0/E_r=1$.}
\label{table2}
\end{table}

\end{document}